%% file: InformationEntropyProduciton.tex
\documentclass[11pt]{article}%{report}
\usepackage{fullpage}
\usepackage[margin=3.3cm]{geometry}
\usepackage{amsmath}
\usepackage{amsfonts}
\usepackage{amssymb}
\usepackage{dsfont}
\usepackage[applemac]{inputenc}
\usepackage[T1]{fontenc}
\usepackage[active]{srcltx}
\usepackage{amsfonts}
\usepackage{amssymb}
\usepackage{amsthm}
\usepackage{graphicx}
\usepackage{color}
\usepackage{bm}% bold math
\usepackage{bbm}
\usepackage{authblk}

\newenvironment{remark}[1][Remark]{\begin{trivlist}
\item[\hskip \labelsep {\bfseries #1}]}{\end{trivlist}}

\input{macro3}

\begin{document}

\title{Information Entropy Production of Spatio-Temporal Maximum Entropy Distributions}
\author[a]{Rodrigo Cofr\'{e}\thanks{rodrigo.cofre@unige.ch}}
\author[b]{Cesar Maldonado}
\affil[a]{Department of Theoretical Physics 
University of Geneva, Switzerland}
\affil[b]{Centro de Modelamiento Matem\'{a}tico. Universidad de Chile. Beauchef 851, Edificio Norte, Piso 7. Santiago, Chile}

\maketitle

\begin{abstract} 
Spiking activity from populations of neurons display causal interactions and memory effects. Therefore, they are expected to show some degree of irreversibility in time. Motivated by the spike train statistics, in this paper we build a framework to quantify the degree of irreversibility of any maximum entropy distribution. Our approach is based on the transfer matrix technique, which enables us to find an homogeneous irreducible Markov chain that shares the same maximum entropy measure. We provide relevant examples in the context of spike train statistics. 

\end{abstract}

\section{Introduction}

Most biological systems (if not all) are out-of-equilibrium systems \cite{palsso:06}, and display some degree of causality i.e, relevant events in the past have an influence on future behavior of the system. A key feature of out-of-equilibrium systems is irreversibility in time.   
Quantifying the degree of irreversibility is, thus, one of the main motivations within the theory of non-equilibrium statistical mechanics.  The quantity that tells us how far a given system is from its equilibrium state\footnote{Here as in statistical physics the word `state' means probability measure.} is called entropy production rate \cite{jiang:04}.

A system in thermodynamic equilibrium is statistically indifferent to a time-reversal action, in the sense that a typical trajectory of the system will look ``equally typical'' if it is run backward in time. This is obviously false when dealing with systems that possesses some kind of causality or, in other words, an arrow-in-time. Hence, a typical trajectory of an irreversible system will be a ``rare'' trajectory if it is reversed in time (with respect to the invariant probability measure of the original process).

In this work we are concerned with causal interactions in populations of spiking neurons. In this context, as some degree of causality is expected, it is natural to ask for the information entropy production of the associated statistical model. The maximum entropy method has become a standard approach to build the probability measure that describes the spike train statistics \cite{schneidman-berry-etal:06,pillow-etal:08,tkacik-etal:13,vasquez-marre-etal:12}. While memory effects could have a non negligible role in the spike train statistics, most studies based on this approach have focused only on synchronous constraints. Considering spatio-temporal constraints requires to include memory within the statistics, so in this case, the dynamics is appropriately described by a discrete time Markov process. As we describe further in this paper, the maximum entropy approach can be extended, within the framework of Markov chains to this case producing Gibbs distributions (in the sense of Bowen) in the spatio-temporal domain.
Therefore the question of measuring the degree of irreversibility of models of spike train statistics that motivates this work, can be re-framed as measuring the entropy production rate of Markov processes generated by maximum entropy distributions.

There is a vast body of theoretical work dealing with fluctuation relations, irreversibility and their relations for Markov chains.  For example, it has been shown that a stationary Markov chain is reversible iff its entropy production vanishes \cite[Chap. 2]{jiang:04}. In~\cite{jiangetal:03} is proved that a for an stationary irreducible finite Markov chain, both with discrete and continuous time parameters, the logarithm of the ratio of the probability distributions of the Markov chain and its time-reversal, converges almost surely to the entropy production rate of the process, which is defined as the relative entropy of the probability distribution of the original process with respect to the reversed one. They also show a large deviation theorem whose rate function has the same symmetry as the Gallavotti-Cohen fluctuation theorem.

The authors of~\cite{maes-redig:00}, discuss the positivity of the entropy production in the context of stochastic systems out of equilibrium. There, they obtained that positivity of  entropy production is equivalent to the violation of the detailed balance in the case of Markov chains and spin-flips processes. Gaspard, in~\cite{gaspard:04}, deduces an expression for the change of entropy as the sum of a quantity called entropy flow plus the entropy production rate for the case of time-discrete Markov chains. Here we will follow his expressions in the case of a Gibbs measure associated to a finite range potential.

Although the notion of the Gibbs distribution extends to processes with infinite memory \cite{fernandez-maillard:05}, and have been used in the context of spike train statistics \cite{cessac-cofre:13,galves:13}, here we focus ourselves on Gibbs distributions associated to finite-range potentials,  because our method is based on the transfer matrix technique. 

The first papers using the maximum entropy method in the field of computational neuroscience~\cite{schneidman-berry-etal:06,pillow-etal:08}, were mainly based on the Ising model. Since then, different objections about their capability to model large networks appeared ~\cite{roudi-nirenberg-etal:09}. Extensions of the Ising model approach have been proposed lately considering the probability of having $K$ neurons firing at the same time bin, the so called $K$-pairwise~\cite{tkacik-etal:13} and models considering spatio-temporal constraints using the Monte Carlo sampling method claimed to correctly fit large sets of spike recordings~\cite{nasser-marre-etal:14}.

This paper is organized as follows: In section 2 we introduce the discrete-time homogeneous Markov chains setup and review the properties we need further. In section 3 we present as pedagogically as possible the transfer matrix technique and its connections with the maximum entropy principle used in spike train statistics.  We introduce in the framework of Gibbs measures the maximum entropy method. We also clarify the role of the potentials and after that we provide the main result of this paper which is the explicit formula to compute the information entropy production solely based on the maximum entropy potentials. In section 4 we provide examples of relevance in the context of spike train statistics. We finish this paper with discussions. 

As this connection is done through Markov chains we can take advantage of results in this field   

\section{Setting}

In order to set a common ground for the analysis of information entropy production of spike trains, here we introduce the notations and provide with the basic definitions used throughout the paper.

\subsection{Notation}

Let us consider a finite network of $N\geq 2$ neurons. We assume that there is a natural time discretization such that at each time step, each neuron emit at most one spike\footnote{There is a minimal amount of time for neurons in which no two spikes can occur called ``refractory period'', when binning, usually one goes beyond this time and eventually two spikes occur at the same time bin. In these cases the convention is to consider only one spike.}. We denote the \textit{spike-state} of each neuron $\sigma_{k}^{n}=1$ whenever the $k$-th neuron emits a spike at time $n$, and $\sigma_{k}^{n}=0$ otherwise. The spike-state of the entire network at time $n$ will be denoted $\sigma^n := \bra{\sigma_k^n}_{k=1}^{N}$, which we call it a \textit{spiking pattern}. For $n_{1}\leq n_{2}$, we use the notation $\sigma^{n_1,n_2}$ for a \textit{spike block}:
\[
\sigma^{n_1,n_2} := \sigma^{n_1}\sigma^{n_1+1}\cdots\sigma^{n_2-1}\sigma^{n_2},
\]
which is an ordered concatenation of spike patterns $\sigma^{n}$. Now, given $T>0$, a \textit{spike train} is a spike block $\sigma^{t_0,T}$. In practice $t_0 =0$. 

\begin{figure}[h!]
  \centering
    \includegraphics[width=0.8\textwidth]{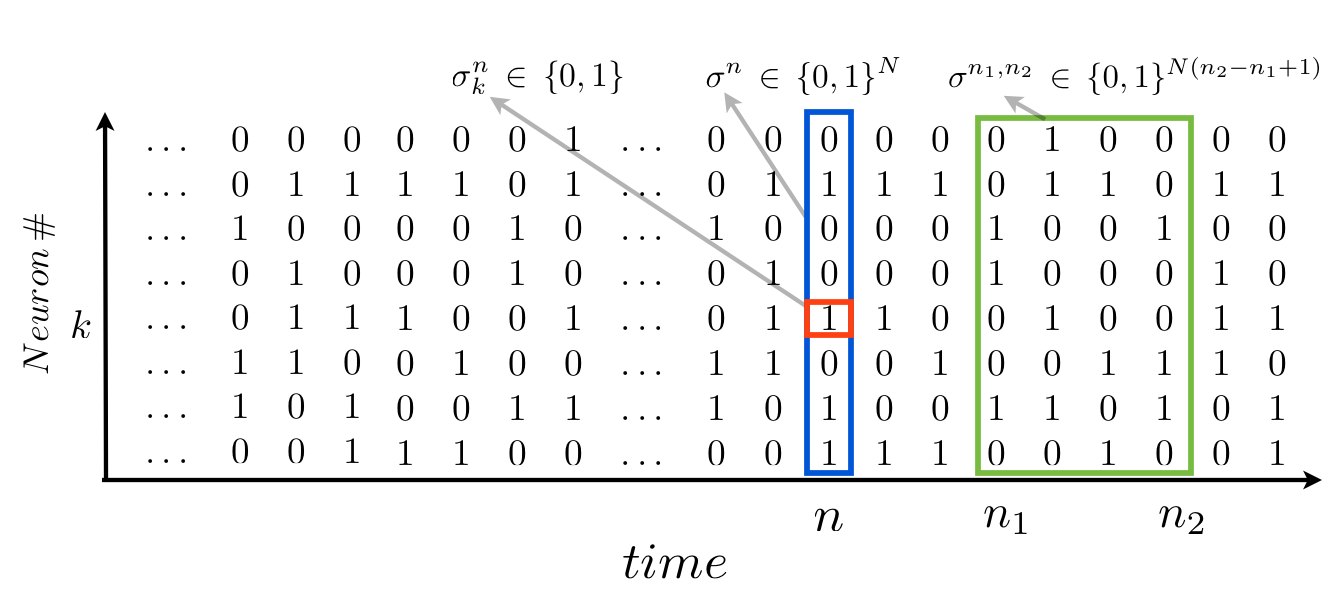}
    \caption{Notation: In red a spike state, in blue a spiking pattern and in green a spike block of length 4.}
\end{figure}

Let $L>0$, we call $\Sigma_N^L := \{0,1\}^{N\times L}$ to the set of spike blocks of $N$ neurons and length $L$. This is the set of $N\times L$ blocks whose entries are $0$'s and $1$'s.
We introduce a symbolic representation to describe the spike blocks. Consider a fixed $N$, then to each spike block $\sigma^{0,L-1}$ we associate a unique number $\ell\in\mathbb{N}$, called \textit{block index}: 

\beq \label{blindex}
\ell(\sigma^{0,L-1}):=\sum_{k=1}^N \sum_{n=0}^{L-1} 2^{n\,N+k-1} \, \sigma_k^n.
\eeq

\noindent
We adopt the following convention: neurons are arranged from bottom to top and time runs from left to right in the spike train. For fixed $N$ and $L$ denote $\sigma^{(\ell)}$ the unique spike block corresponding to the index $\ell$. 
\begin{ex}
Consider $N=2$ and $L=3$, and take $\ell=9$ then the spike block corresponding to the index $\ell$ is
$\sigma^{(9)} \tiny{=\left[
\begin{array}{ccc}
0&1&0\\
1&0&0\\
\end{array}
\right]
}$.
\end{ex}

\begin{ex}
Consider $N=2$ and $L=2$. The state space $\Sigma_2^2$ is given by:
\[
\Sigma_2^2=\Big\{\tiny{\underbrace{\left[
\begin{array}{ccc}
0&0\\
0&0\\
\end{array}
\right]}_{0}
}, \tiny{\underbrace{\left[
\begin{array}{ccc}
0&0\\
1&0\\
\end{array}
\right]}_{1}
},\hdots, \tiny{\underbrace{\left[
\begin{array}{ccc}
1&1\\
1&1\\
\end{array}
\right]}_{15}
}\Big\}.
\]
\end{ex}
\noindent

\subsection{Discrete-time Markov chains and spike train statistics}

Let us consider the random process $\{\xi_{n}:n\geq0\}$ taking values on $\Sigma_{N}^{L}$. In this work, we assume that the spiking activity of the neuronal network can be modeled by some discrete-time Markov process whose transition probabilities are obtained by means of the maximum entropy method. In this setting, $\Sigma_{N}^{L}$ is the state space of the Markov chain, and thus, if $\xi_{n} = \sigma^{n,n+L-1}$ we say that the process is in the state $\sigma^{n,n+L-1}$ at time $n$. The transition probabilities are given as follows, 
\begin{equation}\label{Markov1}
\pp\big[\xi_{n}=\sigma_{(n)} \mid \xi_{n-1}=\sigma_{(n-1)}, \hdots, \xi_{0}=\sigma_{(0)}\big] = \pp\big[\xi_{n}=\sigma_{(n)} \mid \xi_{n-1}=\sigma_{(n-1)}\big],
\end{equation}
where we used the short hand notation $\sigma_{(n)}:= \sigma^{n,n+L-1}$. We emphasize that in this paper the states are spike blocks of finite length $L$, $\sigma^{n,n+L-1}$. 

We further assume that our Markov chain is homogeneous, that is, \eqref{Markov1} is independent of $n$.\\

Since transitions are considered between blocks of the form $\sigma^{n-L,n-1} \rightarrow \sigma^{n-L+1,n}$, therefore the block $\sigma^{n-L+1,n-1}$ must be common for the transition to be possible. Consider two indices $\ell$ and $\ell'$ and their respective spike blocks $\sigma^{(\ell)},\sigma^{(\ell')}\in \Sigma_N^L$ of length $L\geq 2$. We say that the transition $\sigma^{(\ell)} \to \sigma^{(\ell')}$ is \textit{allowed} if $\sigma^{(\ell)}$ and $\sigma^{(\ell')}$ have the common sub-block $\sigma^{1,L-1}= \tilde{\sigma}^{0,L-2}$, where $\tilde{\sigma}^{0,L-2}$ are the first $L-1$ columns of $\sigma^{(\ell')}$.

Now, we define the transition matrix $P:\Sigma_N^L\times\Sigma_N^L\to \setR$, whose entries are given by the transition probabilities, as follows,

\begin{equation}\label{transmatrix}
P_{\ell,\ell'}:= P_{\sigma^{(\ell)},\sigma^{(\ell')}}=
\left\{
\begin{array}{lll}
\pp[\sigma^{(\ell)} \mid \sigma^{(\ell')}] \geq 0 
\quad &\mbox{if }   \sigma^{(\ell)} \to \sigma^{(\ell')}  \mbox{ is allowed } \\
0, \quad &\mbox{otherwise}.
\end{array}
\right.
\end{equation}
\noindent
Note that $P$ has $ 2^{NL} \times 2^{NL}$ entries, but it is a sparse matrix since each line has, at most, $2^N$ non-zero entries.

\begin{ex}
Take $N=2$ and $L=3$, the following represent an allowed transition:
$$
\tiny{\sigma^{(\ell)} =\left[
\begin{array}{ccc}
0&0&1\\
0&1&1\\
\end{array}
\right]
} \rightarrow
\tiny{ \sigma^{(\ell')} =\left[
\begin{array}{ccc}
0&1&1\\
1&1&0\\
\end{array}
\right].
}
$$
Then, the probability to go from $\sigma^{(\ell)}$ to $\sigma^{(\ell')}$ is non-negative,
$$\pp[ 
\tiny{\begin{array}{ccc}
0&1&1\\
1&1&0\\
\end{array}}
\mid 
\tiny{\begin{array}{ccc}
0&0&1\\
0&1&1\\
\end{array}}
 ]=\pp[ 
\tiny{\begin{array}{ccc}
1\\
0\\
\end{array}}
\mid 
\tiny{\begin{array}{ccc}
0&0&1\\
0&1&1\\
\end{array}}
 ] \geq 0.$$
Here, a forbidden transition,
$$
\tiny{\sigma^{(\ell)} =\left[
\begin{array}{ccc}
0&0&1\\
0&1&1\\
\end{array}
\right]
}
\nrightarrow
\tiny{\sigma^{(\ell')} =\left[
\begin{array}{ccc}
0&1&1\\
0&1&0\\
\end{array}
\right]
},
\hspace{0.5cm}
\pp[\tiny{\begin{array}{ccc}
0&1&1\\
0&1&0\\
\end{array}} \mid \tiny{\begin{array}{ccc}
0&0&1\\
0&1&1\\
\end{array}}]=0.
$$
\end{ex}

\begin{figure}[h!]
  \centering
    \includegraphics[width=1.1\textwidth]{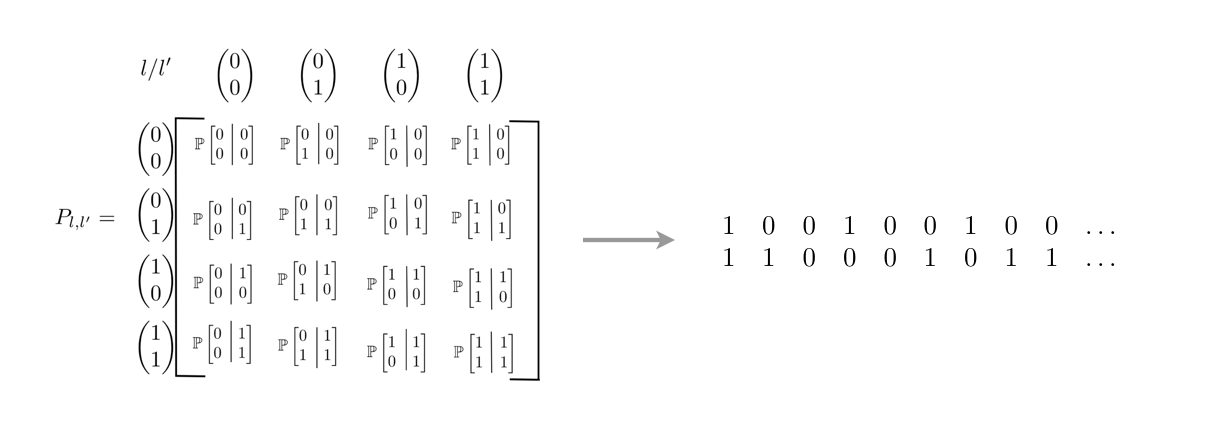}
    \caption{Example of spike train generated by a Markov chain with 4 states, in this example $N=2$ and $L=1$}
    \label{fig:trans}
\end{figure}

\noindent 
A stochastic matrix $P$ is defined from transition probabilities \eqref{transmatrix} satisfying:
\begin{equation*}%\label{sumu}
\mathbb{P}[\sigma^{(\ell)} \mid \sigma^{(\ell')}] \geq 0; \quad \quad \sum_{\sigma^{(\ell')} \in \Sigma} \mathbb{P}[\sigma^{(\ell')} \mid \sigma^{(\ell)}] =1,
\end{equation*}
for all states $\sigma^{(\ell)}, \sigma^{(\ell')}  \in \Sigma_N^L$ (see figure \ref{fig:trans} for an illustration). Moreover, by construction, for any pair of states, there exists a path of maximum length $L$ in the graph of transition probabilities going from one to the other, which means that the Markov chain is primitive. 

\subsection{Homogeneous Markov chain and detailed balance}

Let us call $\sigma^{0,L-1}$ the initial state, and let it be distributed according to some probability measure $\nu$ defined on $\Sigma_N^L$. From the homogeneous Markov property and the definition of the transition matrix, one has 
\begin{equation}\label{Markovdis}
\pp[\xi_{0}=\sigma_{(0)},\xi_{1}=\sigma_{(1)},\hdots,\xi_{k}=\sigma_{(k)}]=\nu(\sigma_{(0)})P_{\sigma_{(0)},\sigma_{(1)}}\cdots P_{\sigma_{(k-1)},\sigma_{(k)}},
\end{equation} 
for all $k>0$, and all states $\sigma_{(0)},\sigma_{(1)},\hdots,\sigma_{(k)}$. Here again, we used the short-hand notation $\sigma_{(k)}\equiv \sigma^{k,L+k-1}$. Equation \eqref{Markovdis} represents the probability law of the homogeneous Markov chain.

The \textit{invariant probability distribution} of the matrix $P$ is a probability vector $\pi$ indexed by the states $\sigma \in \Sigma_N^L$ of the Markov chain, such that
\begin{equation*}%\label{Markovpi2}
\pi^T P=\pi^T,
\end{equation*} 
where $T$ denotes the transpose. 

Let us now consider the general setting of non-stationary distributions. For any initial distribution $\nu$ and for all states $\sigma\in\Sigma_N^L$, one has that
\begin{equation*}%\label{Markovpi3}
\nu(\sigma)= \sum_{\sigma' \in \Sigma_N^L}\nu(\sigma)P_{\sigma,\sigma'}.
\end{equation*} 
Now, let $\nu^{n}$ be the distribution of blocks at time $n$, then one has that the probability evolves in time as follows,
\[
\nu^{n+1}(\sigma)= \sum_{\sigma' \in \Sigma_N^L}\nu^{n}(\sigma')P_{\sigma',\sigma}.
\]
\noindent 
For every $\sigma \in \Sigma_N^L$ one may write the following relation
\begin{equation}\label{evol-proba2}
\nu^{n+1}(\sigma) - \nu^{n}(\sigma) = \sum_{\sigma' \in \Sigma_N^L}\big[ \nu^{n}(\sigma')P_{\sigma',\sigma} - \nu^{n}(\sigma)P_{\sigma,\sigma'} \big].
\end{equation}
This last equation is related to the conditions of reversibility of a Markov chain. When stationarity is assumed, $\nu^{\infty} =\pi$ the stationary distribution is said to satisfy detailed balance if
\begin{equation}\label{det-balance}
\pi_{\sigma}P_{\sigma,\sigma'} = \pi_{\sigma'}P_{\sigma',\sigma} \quad \forall \sigma,\sigma' \in \Sigma_N^L.
\end{equation}
If the detailed balance equations are satisfied, then the quantity inside the parenthesis in the right-hand side of \eqref{evol-proba2} is zero. 

\begin{remark}
\textbf{1:} Detailed balance is a sufficient condition to have a stationary distribution, but it is not necessary.
\end{remark}

\begin{remark}
\textbf{2:} Detailed balance is a common assumption in statistical physics, standard MCMC (Markov chain Monte Carlo) methods are able to estimate (reconstruct) the invariant measure of a Markov process assuming that the detailed balance condition is satisfied.
\end{remark}

\subsection{Information Entropy and information entropy production rates}

A well established measure of the amount of uncertainty of a probability distribution $\nu$ is the \textit{information entropy rate} introduced by Shannon \cite{shannon:48}. We denote it by $\mathcal{S}(\nu)$. In our model, in the case of independent sequences of spike patterns (that is $L=1$), the entropy rate is given by:

\begin{equation*}%\label{entropie_spat}
\mathcal{S}(\nu)=-\sum_{\sigma \in \Sigma_N^1} \nu\bra{\sigma} \log \nu\bra{\sigma}.
\end{equation*}

In a broader setting of stationary  Markov chains in the state space $\Sigma_N^L ; L \geq 2$, and whose invariant distribution is $\pi$, the information entropy rate is given by:
\beq\label{kseq}
\mathcal{S}(\pi) = - \sum_{\sigma, \sigma' \in \Sigma_N^L }\pi(\sigma)P_{\sigma,\sigma'}\log{P_{\sigma,\sigma'}},
;\quad L \geq 2,
\eeq
which in this case corresponds to the Kolmogorov-Sinai entropy \cite{kitchens:98}.

Next, we introduce the information entropy production as in~\cite{gaspard:04}. For expository reasons, let us for the moment consider a non-stationary situation. The information entropy of a probability distribution $\nu$ at time $n$ be given by
\[
\mathcal{S}_{n}(\nu) = -\sum_{\omega \in \Sigma_N^L }\nu^{n}(\omega)\log\nu^{n}(\omega),
\]
so one can consider the \textit{change of entropy} over one time step, as follows
\[
\Delta\mathcal{S}_{n}:= \mathcal{S}_{n+1}(\nu) - \mathcal{S}_{n}(\nu) =   -\sum_{\sigma \in \Sigma_N^L }\nu^{n+1}(\sigma)\log\nu^{n+1}(\sigma)+\sum_{\sigma \in \Sigma_N^L }\nu^{n}(\sigma)\log\nu^{n}(\sigma),
\]
arranging adequately one has~\cite{gaspard:04} that the previous equation can be decomposed as follows
\begin{equation*}
\begin{split}
\Delta\mathcal{S}_{n} = -\sum_{\sigma,\sigma' \in \Sigma_N^L }\nu^{n}(\sigma')P_{\sigma',\sigma}\log\frac{\nu^{n+1}(\sigma')P_{\sigma',\sigma}}{\nu^{n}(\sigma)P_{\sigma,\sigma'}} +  \hspace{2.5cm}\\
 \frac{1}{2}\sum_{\sigma,\sigma' \in \Sigma_N^L }\big[ \nu^{n}(\sigma')P_{\sigma',\sigma}-  \nu^{n}(\sigma)P_{\sigma,\sigma'}\big]\log\frac{\nu^{n}(\sigma')P_{\sigma',\sigma}}{\nu^{n}(\sigma)P_{\sigma,\sigma'}},
\end{split}
\end{equation*}
\noindent
where the first part on the r.h.s is called \textit{information entropy flow} and the second \textit{information entropy production}. 
 
Observe that in the stationary state, one has that $\nu^{n} =\nu^{n+1} = \pi$, thus the change of entropy rate is zero, meaning  information entropy flow equal information entropy production.  

\begin{equation*}
\begin{split}
\Delta\mathcal{S}_{n} = 0= -\sum_{\sigma,\sigma' \in \Sigma_N^L }\pi(\sigma')P_{\sigma',\sigma}\log\frac{\pi(\sigma')P_{\sigma',\sigma}}{\pi(\sigma)P_{\sigma,\sigma'}} +  \hspace{2.5cm}\\
 \frac{1}{2}\sum_{\sigma,\sigma' \in \Sigma_N^L}\big[ \pi(\sigma')P_{\sigma',\sigma}-  \pi(\sigma)P_{\sigma,\sigma'}\big]\log\frac{\pi(\sigma')P_{\sigma',\sigma}}{\pi(\sigma)P_{\sigma,\sigma'}}.
\end{split}
\end{equation*}

In this paper, we focus on the stationary case, in which the entropy production rate will be explicitly given by 
\[
ep(\pi)=\frac{1}{2}\sum_{\sigma,\sigma' \in \Sigma_N^L}\big[ \pi(\sigma')P_{\sigma',\sigma}-  \pi(\sigma)P_{\sigma,\sigma'}\big]\log\frac{\pi(\sigma')P_{\sigma',\sigma}}{\pi(\sigma)P_{\sigma,\sigma'}},
\]
nevertheless, we stress the fact that one can still obtain the information entropy production rate even in the non-stationary case.

The non-negativity implies that information entropy is continuously produced as long as the process has not reached detailed balance. This is analogous to the second law of thermodynamics \cite{nicolis:12}. 
From this equation is easy to realize that if the Markov chain satisfies the detailed balance condition the information entropy production is zero. 

\begin{remark}
\textbf{3:} At first glance it may seem contradictory the fact that in stationary state the information entropy is constant, but at the same time there is a positive ``production'' of information entropy. In stationary state the information entropy production \textit{always} compensate the information entropy flow, thus the information entropy rate remains constant. In this case we refer to non-equilibrium steady states (NESS).  
\end{remark}

\begin{remark}
\textbf{4:} The information entropy production is a functional in the space of probability distributions. For irreducible Markov chains with finite state space is a convex functional, providing a generalized variational principle for non-equilibrium problems \cite{jaynes:80}. The unique minimum entropy production probability measure is often called the \textit{Prigogine distribution.}
\end{remark}

\begin{ex}
Consider $\sigma^{(\ell)}, \sigma^{(\ell')}$ belonging to the state space $\Sigma_N^L$. To alleviate notation we can call $\pi(\sigma^{(\ell)})=\pi_\ell$ and  $P_{\sigma^{(\ell)}, \sigma^{(\ell')}}=P_{\ell,\ell'}$. Consider a spike train generated by a Markov chain in which you know $\pi_\ell$ and $P_{\ell,\ell'} \, \forall \sigma^{(\ell)},\sigma^{(\ell')} \in \Sigma_N^L$. In this case the information entropy production is:

\begin{equation} \label{eq1c}
\begin{split}
ep(\pi) & = \frac{1}{2}\sum_{\sigma^{(\ell)},\sigma^{(\ell')} \in \Sigma_N^L}\Big[\pi_{\ell} P_{\ell,\ell'}-\pi_{\ell} P_{\ell',\ell}\Big]  \log \Big[ \frac{\pi_{\ell} P_{\ell,\ell'}}{\pi_{\ell'}P_{\ell',\ell}} \Big] \\
 & =\sum_{\sigma^{(\ell)},\sigma^{(\ell')} \in \Sigma_N^L}[\pi_{\ell} P_{\ell,\ell'}]  \log \Big[ \frac{\pi_{\ell} P_{\ell,\ell'}}{\pi_{\ell'}P_{\ell',\ell} } \Big]  \geq 0
\end{split}
\end{equation}
\end{ex}

\begin{remark}
\textbf{5:} This last equation emphasizes the fact that information entropy production can also be seen as the Kullback-Leibler divergence between the stochastic process and its time reversal.
\end{remark}

\begin{remark}
\textbf{6:} For spike trains obtained form real recording with more than 10 neurons it is impossible to reliably obtain $\pi_{\ell}$ and $P_{\ell,\ell'}, \, \forall \sigma^{(\ell)},\sigma^{(\ell')}$ without using statistical models. Maximum entropy models provide a convenient method to approach $\pi_{\ell}$ and $P_{\ell,\ell'}$, that is all we need to obtain the information entropy production. This is the main interest of this paper.
\end{remark}

\section{Inferring and Modeling the statistical behavior of spiking neuronal networks}

Consider a network of $N$ neurons in the most simple possible scenario where sequences of spike patterns are considered time independent. The spike patterns $\sigma \in \Sigma_N^1$ can take $2^N$ possible values. The number of neurons $N$ does not need to be very big to arrive to a situation where it is not possible to observe all possible states by doing computer simulations of spiking neurons or doing real data acquisition (2 hours of spike recordings binned at 20 milliseconds produce less than $2^{19}$ spike patterns). Take for instance 100 neurons. The spike pattern can take $2^{100}$ possible values, fact that  makes the frequentist approach unsuitable. The relevant question for the spike train statistics is:  What can we tell about the statistics of these patterns given the fact that we observe a small part of them? In \cite{jaynes:57} the maximum entropy method was introduced, by which one can build a probability distribution uniquely constrained by the available data without the need of counting how many times each pattern appears in the spike train.

\subsection{Maximum Entropy Method}

The maximum entropy approach offers a method for selecting statistical models from first principles. Rooted in statistical mechanics, it consists in solving a constrained maximization problem.  The first step is choosing (arbitrarily) a set of functions (or features) and determine from the data the empirical average these features. These empirical averages are the constraints of the maximization problem. Maximizing the information entropy rate, which is a concave functional in the space of Lagrange multipliers associated to the constraints, provides a unique probability distribution, that turns out to be Gibbs, which approaches at best, the statistics in the following sense: among all probability distributions that matches exactly the constraints, is the one that maximizes the information entropy. The solution satisfies the constraints without adding additional assumptions on the statistics \cite{jaynes:57}. From now on, consider we have a set of spiking data:

$$\mathcal{D}=\{\sigma^0,\sigma^1,\sigma^2,\dots,\sigma^{T-1},\sigma^T\}$$

\begin{remark}
\textbf{7:} Here we consider that $\mathcal{D}$ is the binary representation of the population neural activity. This representation is obtained after a time bin length (time intervals in which spikes are considered synchronous) is fixed. Different choices of time bin length produces different data sets.
\end{remark}

\subsubsection{Observables}

An \textit{observable} is a function, $\cO(\sigma^{0,T})$ that associates a real number to a spike train (spike block). We say that an observable $\cO$ has \textit{range $R$} if it depends on $R$ consecutive spike patterns, e.g. $\cO(\sigma^{0,T})=\cO(\sigma^{0,R-1})$.  We consider here that  observables do not depend explicitly on time (\textit{time-translation invariance of observables}).
As a consequence, for any time $n$, $\cO(\sigma^{0,R-1})=\cO(\sigma^{n,n+R-1})$ whenever $\sigma^{0,R-1}=\sigma^{n,n+R-1}$.
Prominent examples of observables are products of the form:

\beq \label{DefObs}
\cO(\sigma^{0,T})=\prod_{u=1}^r \sigma_{k_u}^{n_u},
\eeq
where $k_u = 1 \dots N$ (neuron index) and $n_u = 0 \dots T$ (time index). These observables are called \textit{monomials} and take values in $\{0,1\}$.  Typical choices of monomials are
$\sigma_{k_1}^{n_1}$ which is $1$ if neuron $k_1$ fires at time $n_1$ and $0$ otherwise; $\sigma_{k_1}^{n_1} \, \sigma_{k_2}^{n_2}$  which is $1$  if neuron $k_1$ fires at time $n_1$ and neuron $k_2$ fires at time $n_2$ and $0$ otherwise. For  $N$ neurons and time range $R$ there are  $2^{NR}$ possible monomials. To alleviate notations, instead of labeling monomials by a list of pairs, as in (\ref{DefObs}), we label them by an integer index, $\ell$ (the index is defined in the same way as the block index (\ref{blindex}), i.e. a monomial reads $m_\ell$. 

\begin{ex}
Consider the state space $\Sigma_3^1$. Here: $m_1=\sigma_1, \, m_2=\sigma_2,\, m_3=\sigma_1\sigma_2,\, m_4=\sigma_3,\, m_5= \sigma_1\sigma_3, \dots$
\end{ex}

\subsubsection{Average of monomials}

As we discuss at the beginning of this section, in general, we cannot obtain samples of conditional probabilities for all the states. However, there is something relatively easy to do when the binary representation of a spike train is available, which is count within the data how many times a neuron has spike or how many times two neurons have fired at the same time, or delayed in time. 
Given a spike train $\mathcal{D}$ of length $T$, we note $\delta_{\mathcal{D}}^{(T)}[\mathcal{O}]$ the empirical average of the observable $\mathcal{O}$. \\

\begin{ex}
Consider a spike train data set $\mathcal{D}$. The empirical firing rate of neuron $k$ is:
\begin{equation*}%\label{Fr}
\delta_{\mathcal{D}}^{(T)} [\sigma_k]=\frac{1}{T} \sum_{n=0}^{T-1}\sigma_k^n;
\end{equation*}
the empirical probability that two neurons $k,j$ fire at the same time is:
\begin{equation*} %\label{Pc}
\delta_{\mathcal{D}}^{(T)} [\sigma_k \sigma_j]=\frac{1}{T}\sum_{n=0}^{T-1}\sigma_k^n \sigma_j^n.
\end{equation*}
and the empirical probability that neuron $k$ fires one time step before neuron $j$:
\begin{equation*}%\label{1t}
\delta_{\mathcal{D}}^{(T)} [\sigma_k \sigma_j^1]=\frac{1}{T-1}\sum_{n=0}^{T-2}\sigma_k^n \sigma_j^{n+1}.
\end{equation*}
\end{ex}
Considering only the empirical average of monomials is not enough to uniquely characterize the spike train statistics. Indeed, there are infinitely many probability distributions sharing the same averages of monomials.

\subsubsection{Potential}

An important example of observable is called \textit{potential}. Potentials of range $R$ can be written as a linear combination of monomials\footnote{The range of the potential is the maximum of the range of the monomials $m_\ell$ considered.}. A potential of range $R$ is written as follows:
\begin{equation*}%\label{H}
\mathcal{H}(\sigma)\, := \, \sum_{\ell=1}^{2^{NR}} h_\ell m_\ell(\sigma) \quad \sigma \in \Sigma_N^R .
\end{equation*}
where the coefficients $h_\ell$ are finite\footnote{Here we do not consider hard core potentials with forbidden configurations.} real numbers. Some coefficients in this expansion may be zero.  We assume throughout this paper that $h_\ell< \infty$.

\subsubsection{Variational principle and maximum entropy distributions}%\label{varpro}

Consider a spike train data set $\mathcal{D}$ and the average value of $K$ observables
$\delta^T_{\mathcal{D}}\bra{\mathcal{O}_k}=C_k, \, k \in \{1,\dots,K\}$. 
We are looking for a probability distribution $\pi$ which maximizes the information entropy among all distributions $\nu$ that matches the expected values of all observables  i.e. $\nu[\mathcal{O}_k]=C_k, \,\forall k \in \{1,\dots,K\}$.  This
is equivalent to solve the following variational problem under constraints:

\begin{equation*}%\label{maxentp}
\mathcal{S}\bra{\pi} = \max \Big\{\mathcal{S}\bra{\nu} : \nu\bra{\mathcal{O}_k} =C_k \quad \forall k \{\in 1,\dots,K \Big\}.
\end{equation*}
As the function $\nu \rightarrow \mathcal{S}\bra{\nu}$ is strictly concave, there is a unique maximizing probability
distribution $\pi$ given the set of values $C_k$. In order to
derive an explicit formula for this $\pi$ we introduce the set of Lagrange multipliers $h_k \in \mathds{R}$ such that the potential $\mathcal{H}=\sum_{k=1}^K h_k\mathcal{O}_k $  and study, the unconstrained problem:

\beq\label{VarPrinc}
\p{\mathcal{H}}=\sup_{\nu \in \cM_{inv}} \Big\{\s{\nu}\, + \, \nu\bra{\mathcal{H}} \Big\} =
\s{\pi}\, + \, \pi\bra{\mathcal{H}},
\eeq
\\
where $\p{\mathcal{H}}$ is called the \textit{free energy or topological pressure}. Its properties allow to perform statistical inference for the maximum entropy problem. $\cM_{inv}$ is the set of invariant measures and  we have $\nu\bra{\mathcal{H}} = \sum_{k=1}^K h_k \, \nu\bra{\cO_k}$ is the average value of
$\mathcal{H}$ with respect to $\nu$, which becomes $\sum_{k=1}^K h_k C_k$. 
The variational principle  (\ref{VarPrinc}) selects, among all possible probabilities $\nu$ the probability $\pi$ realizing the supremum.
In this paper, we only consider potentials $\mathcal{H}$ of finite range. 

In this case the supremum is attained at a unique probability. 
The measure attaining the supremum in this variational principle is well known to satisfy the Gibbs property\footnote{With in the jargon of ergodic theory a measure satisfying the variational principle is called equilibrium state. Under certain circumstances of relevance in this paper equilibrium measures may not satisfy the detailed balance condition (NESS).}.

\subsubsection{Statistical Inference}

The functional $\p{\mathcal{H}}$ has the following properties: Is a log generating function of cumulants:

\beq\label{moy}
\frac{\partial \p{\mathcal{H}}}{\partial h_\ell} =\pi\bra{m_\ell},
\eeq
the average of $m_\ell$ with respect to $\pi$, and:

\beq\label{d2Pbeta}
\frac{\partial^2 \p{\mathcal{H}}}{\partial h_k \partial h_\ell} = \frac{\partial \pi\bra{m_\ell}}{\partial h_k} = \frac{\partial \pi\bra{m_k}}{\partial h_\ell}
=\sum_{n=-\infty}^{+\infty} C_{m_k\,,m_\ell}(n),
\eeq
\[
\mbox{where}\quad C_{m_k\,m_\ell}(n)=\pi\bra{m_k \,m_\ell \circ \rho^n}
 \, - \,  \pi\bra{m_k} \pi\bra{m_\ell},
\]
is the correlation function between the two monomials $m_k$ and $m_\ell$ shifted $n$ time steps ($\rho$ denotes the \textit{shift operator}) in the equilibrium state $\pi$. Correlation functions decay exponentially fast whenever $\mathcal{H}$ has finite range and $\mathcal{H} >-\infty$, thus $\sum_{n=-\infty}^{+\infty} C_{m_k\,,m_\ell}(n) < +\infty$.
Eqn. \eqref{d2Pbeta} characterizes the variation in the average value of $m_\ell$ when varying $h_k$ (linear response).
The corresponding matrix is a \textit{susceptibility matrix}. It controls the  Gaussian fluctuations of monomials around their mean (central limit theorem) \cite{bowen:75}.  When considering potential of range 1, eqn. \eqref{d2Pbeta} reduces to the classical fluctuation-dissipation theorem.
For finite range potentials $\cP(\mathcal{H})$ is a convex function of $h_\ell$'s. This ensures the uniqueness of the solution of \eqref{VarPrinc}.
There exist efficient algorithms to estimate the Lagrange multipliers for the maximum entropy problem with spatio-temporal constraints \cite{nasser-marre-etal:14}.\\

\begin{ex}

The following is a non-exhaustive list of potentials that have been used to fit spike train statistics using the maximum entropy method on spike recordings:

$$\mathcal{H}^1=\sum_{i=1}^N h_i \sigma_i;\quad \mathcal{H}^2= \frac{1}{2}\sum_{i,j=1}^N J_{ij} \sigma_i \, \sigma_j ;\quad \mathcal{H}^3= \frac{1}{3}\sum_{i,j,k=1}^N J'_{ijk} \sigma_i \, \sigma_j\, \sigma_k; $$

$$\mathcal{H}^V=\sum_{k=0}^N \lambda_k \delta(\sum_{i=1}^N  \sigma_i,K) ;\quad \mathcal{H}^t= \sum_{i,j=1}^N \mathsf{\gamma}_{ij} \sigma_i \, \sigma_j^1. $$

\renewcommand{\arraystretch}{1.5}

\begin{center}
    \begin{tabular}{| l | l |  p{3cm}  |}
    \hline
   \textup{Name} &  \textup{Potential} & \textup{Paper}\\ \hline
  Independent & $\mathcal{H}(\sigma) = \mathcal{H}^1$ & \textup{\cite{schneidman-berry-etal:06}}  \\ \hline
  Ising & $\mathcal{H}(\sigma) = \mathcal{H}^1+\mathcal{H}^2$ & \textup{\cite{schneidman-berry-etal:06, roudi-nirenberg-etal:09, shlens-field-etal:06}} \\ \hline
  Triplets & $\mathcal{H}(\sigma) = \mathcal{H}^1+\mathcal{H}^2+\mathcal{H}^3$ & \textup{\cite{ganmor-segev-etal:11a}}\\ \hline
k-Pairwise & $\mathcal{H}(\sigma) = \mathcal{H}^1+\mathcal{H}^2+\mathcal{H}^V$ & \textup{\cite{tkacik-etal:13}}  \\ \hline
Markovian & $\mathcal{H}(\sigma^{0,1})=\mathcal{H}^1+\mathcal{H}^2+\mathcal{H}^t$ & \textup{\cite{marre-etal:09}}  \\ \hline
%\hline
    \end{tabular}

\end{center}
\end{ex}
\noindent
where $\delta$ in $\mathcal{H}^V$ represents the Kronecker delta. These examples highlight the idea of \textit{choice} underlying the maximum entropy distributions. The arbitrary choice of constraints fix the shape of the potential. The task of maximum entropy method is to fit the parameters $\{h_i,J_{ij},J^1_{ij},\gamma_{ij}, \lambda_k\}$ in such a way that the associated probability measure maximizes the information entropy among those that satisfy the constraints. These parameters are the Lagrange multipliers of the constrained maximization problem. 

\subsubsection{Transfer Matrix}%\label{TransfM}

Consider a finite state space $\Sigma_N^L$ and a potential function $\mathcal{H}(\sigma^{0,L-1})$. For $L \geq 2$, each block $\sigma^{0,L-1}$ can be viewed as composed by an allowed transition $\sigma^{(\ell)}=\sigma^{0,L-2} \to \sigma^{(\ell')}=\sigma^{1,L-1} \in \Sigma_N^{L-1}$, in this case we write  $\sigma^{0,L-1} \sim \sigma^{(\ell)}, \sigma^{(\ell')}$. 

In an analogous way as done for Markov approximations of Gibbs measures~\cite{chazottesetal:05, msalgado:13}, we introduce here the \textit{transfer matrix} $\cL$, which in some sense is a generalization of the concept of Markov transition matrix, as follows: 

\beq\label{transfmatrix}
\cL_{\ell,\ell'}=   
\left\{
\begin{array}{lll}
 e^{\mathcal{H}(\sigma^{0,L-1})}
\quad &\mbox{if }  \sigma^{0,L-1} \sim \sigma^{(\ell)}\sigma^{(\ell')}   \\
0, \quad &\mbox{otherwise}.
\end{array}
\right. 
\eeq
From the assumption $\mathcal{H} > -\infty$, each allowed transition corresponds to a positive entry in the matrix $\cL$. As $\cL$ is a primitive matrix, it satisfies the Perron-Frobenius theorem \cite{gantmacher:66}. Let $s > 0$ be its spectral radius. Because of the irreducibility of $\mathcal{L}, s$ is an eigenvalue of multiplicity 1 strictly larger in modulus than the other eigenvalues. Let us call $[L(\sigma^{(\ell)})\equiv L_\ell]_{\sigma^{(\ell)} \in \Sigma_N^{L-1}}$ and $[R(\sigma^{(\ell)})\equiv R_\ell]_{\sigma^{(\ell)} \in \Sigma_N^{L-1}}$ be the left and right eigenvectors of $\mathcal{L}_{\ell,\ell'}$  corresponding to the eigenvalue $s$,

\begin{align*} 
\sum_{\sigma^{(\ell)}} L_\ell \, \cL_{\ell,\ell'} &=s \, L_{\ell'}, \quad \forall \sigma^{(\ell')} \in \Sigma_N^{L-1},   \\ 
\sum_{\sigma^{(\ell')} }  \cL_{\ell,\ell'} \, R_\ell &=s \, R_\ell,  \quad \forall \sigma^{(\ell)} \in \Sigma_N^{L-1},
\end{align*}
Notice that $L_\ell>0$ and $R_\ell>0$ for all $\sigma^{(\ell)} \in \Sigma_N^{L-1}$. In order to obtain the transition matrix $P_{\ell,\ell'}$ corresponding to the associated Markov chain we use the notion of \textit{physically equivalent potentials}. That is, a set of potentials that share the same associated measure (Gibbs and in this context, of maximal entropy), using the so called cohomology equation \cite{walters:75}:

\beq\label{cohe}
\phi_{\ell,\ell'}:=\log P_{\ell,\ell'} = \mathcal{H}_{\ell,\ell'} -\log[R_\ell]+\log[R_{\ell'}] - \log[s].
\eeq
\noindent
This transformation define an irreducible homogeneous Markov chain with transition probability $e^{\phi( \sigma^{0,L-1})}=\pp[\sigma^{1,L-1}\mid \sigma^{0,L-2}]$, from a potential $\mathcal{H}$ and the associated transfer matrix. Taking exponential in~\eqref{cohe} we get $P_{\ell,\ell'}$.

\beq\label{tpmep}
P_{\ell,\ell'}:=\frac{e^{\mathcal{H}_{\ell,\ell'}} R_{\ell'}}{R_\ell \, s}, \quad \forall \sigma^{(\ell)},\sigma^{(\ell')} \in \Sigma_N^{L-1}.
\eeq

\noindent
The unique stationary probability measure is given by: 
\beq\label{ssmep}
\pi_\ell:=\frac{L_\ell \, R_\ell}{\langle L,R \rangle}, \quad \forall \sigma^{(\ell)} \in \Sigma_N^{L-1}.
\eeq

\noindent
It follows from the Markov property and from (\ref{tpmep},\ref{ssmep}) that we can obtain the probability of any block of length $n>L$:
\beq\label{fchc}
\pi[\sigma^{0,n}] = \lpfc{ \sigma^{0,L-2}}\frac{e^{\sum_{k=0}^{n-L+1}\mathcal{H}\pare{\sigma^{k,k+L-1}}}}{s^{n-L+2}} 
\rpfc{\sigma^{n-L+2,n}}.
\eeq

\noindent
This can be easily verified  writing the Markov property using the normalized potential: 

\beq\label{chk-no}
\pi[\sigma^{0,n}]=e^{\sum_{k=0}^{n-L+1} \phi(\sigma^{k,k+L-1})}\pi[\sigma^{0,{L-2}}].
\eeq
\\
Using equation (\ref{cohe}), we have:
\begin{align*}
\sum_{k=0}^{n-L+1} \phi(\sigma^{k,k+L-1}) & = 
\sum_{k=0}^{n-L+1} \Big(\mathcal{H}(\sigma^{k,k+L-1}) -\log R(\sigma^{k,k+L-2}) + \log
R(\sigma^{k+1,k+L-1}) -\log s \Big)      \nonumber \\
 & =   \sum_{k=0}^{n-L+1} \mathcal{H}(\sigma^{k,k+L-1}) -\log
R(\sigma^{0,L-2}) + \log R(\sigma^{n-L+2,n}) -(n-L+2)\log s.
\end{align*}
\noindent
Where the last equality is obtained applying the sum to each term (most of the terms $\log R(\cdot)$ cancel out). Taking exponential we get:
\[
e^{\sum_{k=0}^{n-L+1} \phi(\sigma^{k,k+L-1})} = e^{\Big(
\sum_{k=0}^{n-L+1} \mathcal{H}(\sigma^{k,k+L-1}) -\log
R(\sigma^{0,L-2}) + \log R(\sigma^{n-L+2,n}) -(n-L+2)\log s \Big)},
\]
using properties of the exponential function and multiplying both sides by (\ref{ssmep}) we get:
\[
e^{\sum_{k=0}^{n-L+1} \phi(\sigma^{k,k+L-1})} \pi[\sigma^{0,L-2}] =
L(\sigma^{0,L-2})\frac{e^{\sum_{k=0}^{n-L+1} \mathcal{H}(\sigma^{k,k+L-1})}
R(\sigma^{0,L-2})}{R(\sigma^{0,L-2})s^{n-L+2}}R(\sigma^{n-L+2,n}).
\]
\noindent
Finally, in the l.h.s we use equation (\ref{chk-no}) and the r.h.s is obtained canceling out the term $R(\sigma^{0,L-2})>0$.

The invariant measure of the Markov chain $\pi$ obeys the variational principle (\ref{VarPrinc}) and:

\begin{equation*}%\label{logs}
\mathcal{P}[\mathcal{H}] =\log s.
\end{equation*}

\noindent
Providing a direct way to compute the maximum entropy coefficients (see the next section for an explicit example). When considering a normalized potential $\phi$, the transfer matrix becomes a stochastic transition matrix with maximal eigenvalue 1. Thus $\mathcal{P}[\phi]=0.$ 

In summary, in order to characterize the spike train statistics of a neuronal network whose spikes are interacting through a potential $\mathcal{H}$, the transfer matrix representation and the classical results of non-negative matrices is all that is needed.

It follows from \eqref{fchc} that there exist constants $A, B >0$ such that, for any block $\sigma^{0,n}$ the invariant distribution obeys \cite{bowen:75}:
\beq\label{gibbs2}
A \leq \frac{\pi\bra{\sigma^{0,n}}}{e^{-(n-L+2)\cP(\mathcal{H})} e^{-\sum_{k=0}^{n-L+1}\mathcal{H}\pare{\sigma^{k,k+L-1}}}}\leq B.
\eeq

\noindent
This definition encompasses the classical definition of Gibbs distributions, $\frac{e^{\mathcal{H}}}{Z}$ found in standard textbooks of statistical physics.
%Note that from equation (\ref{fchc}): $\pi[\sigma^{0,n}]$ has not the form $\frac{e^{\mathcal{H}(\sigma)}}{Z_n}$ with $Z_n=\sum_{\sigma^{0,n}} e^{\mathcal{H}(\sigma^{0,n})}$. 
This measure is often called in ergodic theory \textit{``Gibbs in the sense of Bowen''} \cite{chazz:03}.

\begin{remark}
\textbf{8:} Here we have obtained the transfer matrix and its unique invariant probability measure from a potential of finite range, which is a Markov measure that satisfies the variational principle~\cite{chazottesetal:05}. Moreover, is the same as the unique Gibbs measure associated to a finite range potential, that  takes the form \eqref{fchc}~\cite{msalgado:13}. For finite range potentials $\mathcal{H}$ in one dimension, one has always a unique equilibrium measure which satisfies the ``Gibbs property'' \eqref{gibbs2}.  

In other words, what we have built here, is the Gibbs measure associated to the finite range potential $\mathcal{H}$ that by construction is a maximum entropy measure. 
\end{remark}

\subsection{Information entropy production in maximum entropy distributions}

Consider a state space $\Sigma_{N}^{L-1}$ and a fixed potential $\mathcal{H}$. From the results of the previous section, one can uniquely construct a transfer matrix $\mathcal{L}_{\mathcal{H}}$ and thus by means of equation~\eqref{cohe}, the stochastic matrix $P$. Furthermore, one can plug~\eqref{tpmep} and \eqref{ssmep} into the equation~\eqref{eq1c}, providing in this way a formula for the information entropy production that depends only on the transfer matrix $\mathcal{L}_{\mathcal{H}}$. After simplifying we obtain:

\beq\label{epmep}
ep(\mathcal{L}_{\mathcal{H}})=\sum_{\sigma^{(\ell)}, \sigma^{(\ell')} \in \Sigma_N^{L-1}}\frac{L_\ell}{\langle L,R\rangle}\frac{e^{\mathcal{H}_{\ell,\ell'}}R_{\ell'}}{s} \log \Big[ \frac{L_\ell \, R_{\ell'}\, e^{\mathcal{H}_{\ell,\ell'}}}{L_{\ell'}\, R_{\ell'}e^{\mathcal{H}_{\ell',\ell}}} \Big].
\eeq

\noindent
This is a quantity of major interest since from it one is able to measure the degree of irreversibility of the Markov process characterized by the  potential $\mathcal{H}$.

\subsubsection{Conditions for detailed balance in maximum entropy potentials}

Again we can apply \eqref{tpmep} and \eqref{ssmep} to equation \eqref{det-balance}, and we obtain:

$$\frac{L_\ell \, R_\ell }{\langle L,R\rangle}
\frac{ e^{\mathcal{H}_{\ell,\ell'}}\,R_{\ell'}}{R_\ell \, s} = \frac{L_{\ell'} \, R_{\ell'}}{\langle L,R\rangle}
\frac{ e^{\mathcal{H}_{\ell',\ell}} \, R_\ell }{R_{\ell'} \,  s}.$$
Simplifying we get:

\begin{equation*}%\label{dbpf}
\frac{e^{\mathcal{H}_{\ell,\ell'}}}{e^{\mathcal{H}_{\ell',\ell}}} = \frac{R_\ell \, L_{\ell'}}{R_{\ell'} \,  L_\ell}.
\end{equation*}

\noindent
Which is a condition for detailed balance only based on the transfer matrix obtained from the  potential $\mathcal{H}$.

\section{Examples}

In this section we give examples of application of our results.  We detail the transfer matrix technique to compute the Markov transition matrix and its invariant measure from a potential $\mathcal{H}$.  

\subsection{First example: Toy model}

Let us consider a range-$2$ potential with two neurons:

$$\mathcal{H}(\sigma^{0,1})=h_1 \sigma_1^1 \sigma_2^0.$$
\noindent
The transfer matrix (\ref{transfmatrix}) associated to $\mathcal{H}$ is in this case a $4 \times 4$ matrix, as in figure \ref{fig:trans}.
$$
\cL_{\sigma^0,\sigma^1}=
\pare{
\begin{array}{ccccc}
1 & 1 & 1 & 1\\
1 & 1 & 1 & 1\\
1 & e^{h_1}  & 1 & e^{h_1} \\
1 & e^{h_1}  & 1 & e^{h_1} 
\end{array}
}.
$$
\noindent
As this matrix is primitive by construction, it satisfies the hypothesis of the Perron-Frobenius theorem. Its unique maximum eigenvalue is $s=e^{h_1}+3$. The left and right eigenvectors associated to this largest eigenvalue are respectively:
$$\lpfc{\begin{array}{ccc}
0\\
0\\
\end{array}}= \frac{2}{1+e^{h_1}}; \ \lpfc{\begin{array}{ccc}
0\\
1\\
\end{array}}= 1;\ \lpfc{\begin{array}{ccc}
1\\
0\\
\end{array}}=  \frac{2}{1+e^{h_1}};\ \lpfc{\begin{array}{ccc}
1\\
1\\
\end{array}}= 1,$$
$$\rpfc{\begin{array}{ccc}
0\\
0\\
\end{array}}=  \frac{2}{1+e^{h_1}};\ \rpfc{\begin{array}{ccc}
0\\
1\\
\end{array}}=  \frac{2}{1+e^{h_1}};\ \rpfc{\begin{array}{ccc}
1\\
0\\
\end{array}}= 1;\ \rpfc{\begin{array}{ccc}
1\\
1\\
\end{array}}= 1.$$

\noindent
Taking the exponential of equation ~\eqref{cohe} we obtain the transition matrix \eqref{transmatrix}, which reads,
$$
P_{\sigma^0,\sigma^1}=e^{\phi(\sigma^{0,1})}=
\frac{1}{s}
\,
\pare{
\begin{array}{ccccc}
1 & 1 & \frac{1}{r_0} & \frac{1}{r_0}\\
1 & 1 & \frac{1}{r_0} & \frac{1}{r_0}\\
r_0 & e^{h_1}r_0 & 1 & e^{h_1}\\
r_0 & e^{h_1}r_0 & 1 & e^{h_1}
\end{array}
},
$$
\noindent
where $\rpfc{\begin{array}{ccc}
0\\
0\\
\end{array}}=  \frac{2}{1+e^{h_1}}=r_0$. Note that $e^\phi$ is a stochastic matrix (the sum of entries on each row is equal to $1$).

The unique invariant probability of this irreducible Markov chain is given by equation \eqref{ssmep}, and its entries are given by,
$$\pi \Big(\begin{array}{ccc}
0\\
0\\
\end{array}\Big) =\frac{4}{s^2}, \quad \pi\Big(\begin{array}{ccc}
0\\
1\\
\end{array}\Big)=\frac{2(s-2)}{s^2}, \quad \pi\Big(\begin{array}{ccc}
1\\
0\\
\end{array}\Big)=\frac{2(s-2)}{s^2}, \quad \pi\Big(\begin{array}{ccc}
1\\
1\\
\end{array} \Big)=\frac{(s-2)^2}{s^2}.$$
It is easy to check that probability distribution $\pi$ is invariant w.r.t. the transition matrix $P=e^{\phi}$, that is $\pi^{T} P=\pi^{T}$.

With these equations, we can verify that \textit{in general the detailed balance condition is not satisfied}; for example:
$$ P\Big(\begin{array}{ccc}
0\\
1\\
\end{array}     \biggr \rvert  \begin{array}{ccc}
1\\
0\\
\end{array}  \Big) \pi\Big(\begin{array}{ccc}
1\\
0\\
\end{array} \Big) \neq  P\Big(\begin{array}{ccc}
1\\
0\\
\end{array}     \biggr \rvert  \begin{array}{ccc}
0\\
1\\
\end{array}  \Big) \pi\Big(\begin{array}{ccc}
0\\
1\\
\end{array} \Big). $$

\noindent
As we can see in figure \ref{fig:epkse}, the maximum entropy distribution for the unconstrained problem considered so far is attained at $h_1=0$, and is the uniform distribution as expected.

Let us now consider a constrained version of this problem. Suppose we are given with a data set $\mathcal{D}$ and we measure from data the following restriction:
$$\delta_{\mathcal{D}}^T[\sigma_1^1\sigma_2^0]=0.1$$
\noindent
meaning that in the data we found that neuron 1 fires 1 time step after neuron 2 $10\%$ of the time. Given this restriction and using the equation \eqref{moy}, we obtain that,
$$ \frac{\partial \log(e^{h_1}+3)}{\partial h_1} =0.1. $$
\noindent
Solving this equation we find $h_1=-1.09861$. That is among all the distributions that match exactly the restriction, the one that maximizes the information entropy is the one obtained by fixing $h_1$ at the found value, which in this case, is a non-equilibrium steady state (see fig. \ref{fig:epkse}). Is easy to check that the variational principle \eqref{VarPrinc} is satisfied.

\begin{remark}
\textbf{9:} When the product of time range of the potential and the number of neurons is less than $\sim 10$ and with enough data, it is still possible to approximate the invariant measure just counting how many times the process visits a given element in the state space, so the maximum entropy method is in principle unnecessary.
\end{remark}

\subsubsection{Information Entropy Production for this example}

\begin{figure}[h!]
  \centering
    \includegraphics[width=0.8\textwidth]{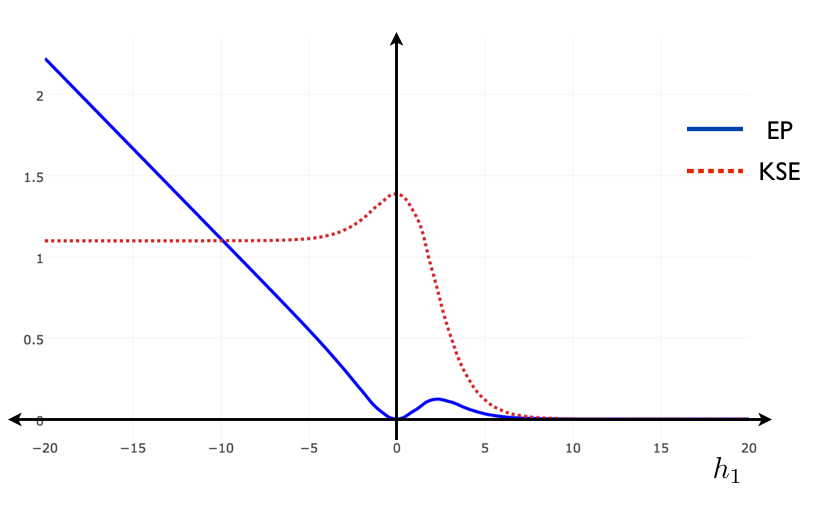}
    \caption{Information entropy production and Kolmogorov-Sinai entropy as a function of $h_1$. Note that this is the unconstrained maximum entropy problem, thus as expected the maximum is attained at $h_1=0$ for the uniform distribution, which is also the Prigogine distribution of minimal information entropy production.}
    \label{fig:epkse}
\end{figure}

Having the transition probability matrix $P$ and the invariant measure $\pi$, we can compute the information entropy rate \eqref{kseq} and the information entropy production \eqref{epmep} as a function of the parameter $h_1$ (see figure \ref{fig:epkse}).

\begin{remark}
\textbf{10:} In this example the detailed balance condition is only satisfied in the trivial case $h_1=0$, in this case the maximum eigenvalue $s=4$ and the invariant probability is the uniform assigning $\frac{1}{4}$ to each spike pattern.
\end{remark}

\subsection{Second example: Memoryless potentials}

Consider a potential of range $R=1$. This case include the Ising model, Triplets, $K$-pairwise and all other memoryless potentials, in the context of neural network models. It represent a limit case in the definition of the transfer matrix where transitions between spike patterns $\sigma' \rightarrow \sigma; \, \sigma,\sigma' \in \Sigma_N^1$ are considered and where all transitions are allowed, but here the potential do not ``see'' the past i.e. $\mathcal{L}_{\sigma',\sigma} = e^{\mathcal{H}(\sigma)}$, thus each column of this transfer matrix has the form:
$$(e^{\mathcal{H}(\sigma)},e^{\mathcal{H}(\sigma)},\hdots,e^{\mathcal{H}(\sigma)}).$$
The matrix $\mathcal{L}$ is degenerated with maximum eigenvalue:
$$s=Z:=\sum_{\sigma} e^{\mathcal{H}(\sigma)}$$
and all other eigenvalues $0$. The left and right eigenvectors corresponding to $s=Z$ are:
$$L(\sigma)=\frac{1}{Z}, \quad R(\sigma)=e^{\mathcal{H}(\sigma)}; \quad \forall\ \sigma \in \Sigma_N^1.$$
Note that $\langle L,R \rangle =1$. We have therefore:
\begin{equation*}%\label{gibbs}
P_{\sigma',\sigma}=P(\sigma) = \pi(\sigma)=\frac{e^{\mathcal{H}(\sigma)}}{Z}; \quad \forall\ \sigma,\sigma' \in \Sigma_N^1,
\end{equation*}
\noindent
In this case the invariant measure $\pi$ has the classical form for the Gibbs distribution. The associated Markov chain has no memory: successive events are independent. This last remark reflects a central weakness of memory-less maximum entropy models to describe neuron dynamics. They neither involve memory nor time causality.

\subsubsection{Information Entropy Production for this example}

Taking the formula of information entropy production \eqref{epmep} we obtain:

$$ep(\mathcal{L}_{\mathcal{H}})=\sum_{\sigma, \sigma' \in \Sigma_N^1}\frac{L(\sigma)}{\langle L,R\rangle}\frac{e^{\mathcal{H}(\sigma')}R(\sigma')}{\log(Z)} \Big(\mathcal{H}(\sigma')-\mathcal{H}(\sigma)\Big) =0.$$
\begin{flushleft}
\textbf{Interpretation}
\end{flushleft}
In the case where only range 1 observables are chosen (firing rates, pairwise correlations, triplets, etc.), the average value of these observables in a given data set is exactly the same as the one taken from another data set where the time indexes have been randomly shuffled or even time inverted. As this is the only information about the process that the maximum entropy method consider, it is not surprising that the stochastic process associated to the maximum entropy distribution is, in this case, time  reversible. Consider a data set $\mathcal{D}$ and a function $g:\{0,\dots,T\} \rightarrow \{0,\dots,T\}$ that randomly shuffles the time indexes. We call the new data set formed by this transformation $\mathcal{D}^{RS}$. We call the data set formed by inverting the time indexes $\mathcal{D}^I$.
\begin{align*} 
\mathcal{D} &=  \{\sigma^0,\sigma^1,\sigma^2,\dots,\sigma^{T-1},\sigma^T\} \\ 
\mathcal{D}^{RS} &=  \{\sigma^{g(0)},\sigma^{g(1)},\sigma^{g(2)},\dots,\sigma^{{g(T-1)}},\sigma^{g(T)}\} \\ 
\mathcal{D}^I &=  \{\sigma^T,\sigma^{T-1},\sigma^{T-2},\dots,\sigma^1,\sigma^0\}.
\end{align*}

\noindent
In these 3 cases which may correspond to very different biological experiments, the average value of every range 1 observable is exactly the same, therefore these data sets are characterized by the same maximum entropy distribution. In particular the stochastic process generating the spike train has the same statistical behavior as their time inverted version, i.e. 

$$\delta_{\mathcal{D}^O}^{(T)} [m_\ell]=\delta_{\mathcal{D}^{RS}}^{(T)} [m_\ell]=\delta_{\mathcal{D}^I}^{(T)} [m_\ell], \quad \forall\ m_l \mbox{ purely spatial.}$$

\noindent
In particular, the firing rates:

$$\delta_{\mathcal{D}^O}^{(T)} [\sigma_k]=\delta_{\mathcal{D}^{RS}}^{(T)} [\sigma_k]=\delta_{\mathcal{D}^I}^{(T)} [\sigma_k], \quad \forall\ k \in \{1,\dots,N\}$$
and pairwise correlations:

 $$\delta_{\mathcal{D}^O}^{(T)} [\sigma_k\sigma_j]=\delta_{\mathcal{D}^{RS}}^{(T)} [\sigma_k\sigma_j]=\delta_{\mathcal{D}^I}^{(T)} [\sigma_k\sigma_j], \quad \forall\ k,j \in \{1,\dots,N\}.$$

\subsection{Third example: 1-time step Markov}

\begin{figure}[h!]
  \centering
    \includegraphics[width=0.99\textwidth]{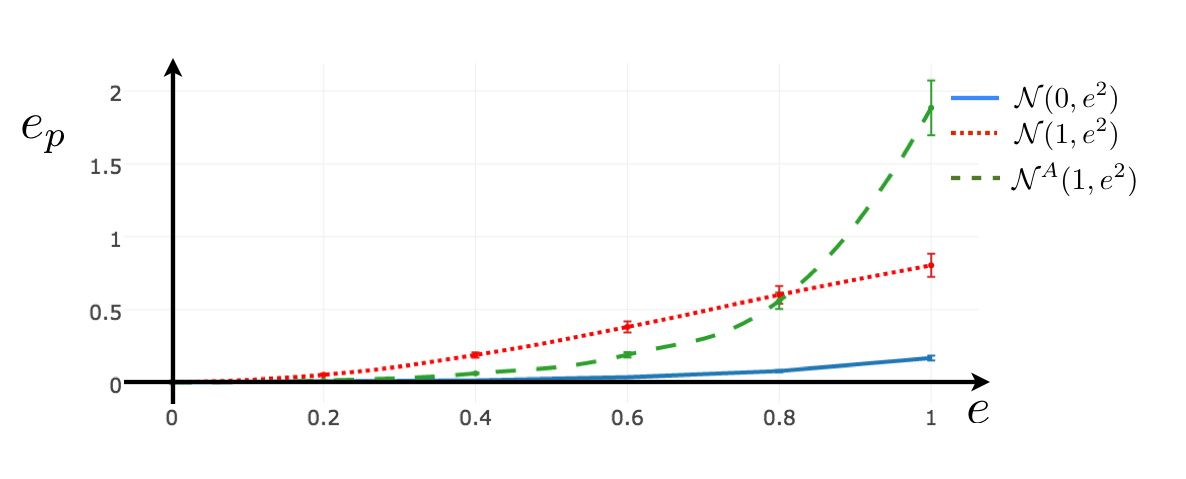}
    \caption{Information entropy production for the 1-time step Markov potential. The parameters $h_i$ and $J_{ij}$ are draw at random one time and remain fixed. We take 100 realizations of the matrix $\gamma_{ij}$ at random but from a Gaussian distribution with different values of mean and standard deviation $e$. $\mathcal{N}^A$ antisymmetric random matrix. We plot the average value of information entropy production for each case, with the respective error bars. }
    \label{fig:ex3}
\end{figure}

Here, we consider the 1-time step extension of the Ising model, that reads:
\beq\label{pm1}
\mathcal{H}(\sigma^{0,1}) = \sum_{i=1}^N h_i \sigma_i +\frac{1}{2} \sum_{i,j=1}^N  J_{ij} \sigma_i \, \sigma_j+ \sum_{i,j=1}^N \mathsf{\gamma}_{ij} \sigma_i \, \sigma_j^1.
\eeq
\\
This is the potential considered to fit a maximum entropy distribution to spiking data from a mammalian parietal cortex in vivo \cite{marre-etal:09}. It is important to notice that in their study, Marre \textit{et al.} compute the solution of the maximum entropy problem using Monte Carlo simulations imposing detailed balance condition, so in their case,  the information entropy production is zero by construction. Here, we do not consider any data set, we rather investigate the capability of this potential to generate information entropy production by considering the following scenarios: We consider a network of $N=10$ neurons, where we draw at random the coefficients $h_i$ and $J_{ij}$ in a range plausible to be the maximum entropy coefficients (or Lagrange multipliers) of an experiment of retinal ganglion cells exposed to natural stimuli  (values of from $h_i$ and $J_{ij}$ as in  \cite{tkacik:15}). We generate the matrix of components $\gamma_{ij}$ by drawing each component at random from a Gaussian distribution. We  summarize our results in figure \ref{fig:ex3}. We observe the following: Independent of  $h_i$ and $J_{ij}$ and the parameters of mean and variance from which the matrix of coefficients $\gamma_{ij}$ are generated, if the matrix of components $\gamma_{ij}$ is symmetric the Markov  process generated by the potential \eqref{pm1} is reversible in time, so the information entropy production is zero. This includes the limit case  when  $\gamma_{ij}=0 , \forall i,j \in \{1,\dots,N\}$, where we recover the Ising model. Next, we fix the values of $h_i$ and $J_{ij}$ (random values), and we generate 100 matrices $\gamma_{ij}$ by drawing their components from Gaussian distributions $\mathcal{N}(0,e^2)$, another 100 from $\mathcal{N}(1,e^2)$. We also generate 100 anti-symmetric matrix  $\gamma_{ij}$ from $\mathcal{N}(1,e^2)$, that we denote in figure \ref{fig:ex3} $\mathcal{N}^A(1,e^2)$. For each realization of the matrix $\gamma_{ij}$ we generate the transfer matrix and proceed as explained in section (3) to obtain the entropy production in each case.  We plot in figure \ref{fig:ex3} the average value of the information entropy production and error bars for each case.

\section{Conclusion} 

We have presented a method to compute the information entropy production of any maximum entropy potential with an emphasis on spike train statistics. We have shown
that information entropy production, which is a non-negative
quantity, takes value 0 for time independent processes (time-reversible) derived in this context from range 1 potentials, for example: Ising, $K$-pairwise, triplets, among others. Spatio-temporal constraints in the context of maximum entropy method, produce homogeneous irreducible Markov chains whose unique steady state is in general of non-equilibrium (NESS), thus detailed balance condition is not satisfied causing strictly positive entropy production. This fact highlights an important issue, \textit{only} spatio-temporal maximum entropy models induce time irreversible processes, feature expected from biological systems. 
We have also identified limiting conditions in the spatio-temporal maximum entropy parameters, under which the information entropy production is strictly positive. 

Our results can be applied also to neuronal network models like integrate and fire and Generalized Linear Models driven by noise, but without time-dependent external stimulus, using their maximum entropy version (see \cite{cofre-cessac:14} for details about how to obtain a maximum entropy potential from a neuronal network model).

There are two main drawbacks of our approach. The first is inherited from the maximum entropy method which requires stationarity in the data. Nevertheless, information entropy production is a much broader concept which can also be measured along non-stationary trajectories.  The second is that is based on the transfer matrix technique, so it requires an important computational effort for large-scale neural networks. 

We believe that there is a lot of room to progress concerning this work, one possibility is to measure the information entropy production for different choices of spatio-temporal constraints using the maximum entropy method on biological spike train recordings. In particular, would be interesting to measure this quantity for retinal ganglion cells responding to different stimulus scenarios. A more ambitious goal would be to link the information entropy production as a signature of an underlying physiological process such as adaptation or learning.  Another direction of interest is to measure the information entropy production for time dependent models where transition probabilities are explicitly given or can be computed \cite{ mora:15, roudi-hertz:11}. Concerning time dependent neuronal network models, future studies will lead to a better understanding of the role and impact of synaptic topology connectivity in neuronal network models or other parameters defining the model in the information entropy production. Previous studies in spike train statistics have measure the dynamical entropy production in spiking neuronal networks using a deterministic approach and based on the Pesin identity (sum of positive Lyapunov exponents)  \cite{monteforte-wolf:10}. There are relationships between the deterministic and stochastic dynamics \cite{gaspard:07}, and some interpretations of deterministic dynamical entropy production with information loss. It would be interesting to investigate whether this relationships bring new knowledge in the field of computational neuroscience. 

While we have mainly focused in this paper on spike train statistics, our results are not restricted to this application and can be used in any field where maximum entropy potentials are considered such as ecology, image processing, economy among others. \\

\textbf{Acknowledgments.} We thank J.-P. Eckmann and Fernando Rosas for discussions and careful
reading of the manuscript. RC was supported by an ERC advanced grant ``Bridges''. CM was supported by the CONICYT-FONDECYT Postdoctoral Grant No. 3140572.

\end{document}

%% file: macro3.tex
\newcommand\seq[3]{{#1}_{#2}^{#3}}
\newcommand{\bd}{\begin{document}}
\newcommand{\ed}{\end{document}}
\newcommand{\beq}{\begin{equation}}
\newcommand{\eeq}{\end{equation}}
\newcommand{\bef}{\begin{figure}}
\newcommand{\enf}{\end{figure}}
\newcommand{\bea}{\begin{eqnarray}}
\newcommand{\eea}{\end{eqnarray}}
\newcommand{\baR}{\begin{array}}
\newcommand{\eaR}{\end{array}}
\newcommand{\bc}{\begin{center}}
\newcommand{\ec}{\end{center}}
\newcommand{\ben}{\begin{enumerate}}
\newcommand{\een}{\end{enumerate}}
\newcommand{\bit}{\begin{itemize}}
\newcommand{\eit}{\end{itemize}}
\newcommand{\su}{\section}
\newcommand{\ssu}{\subsection}
\newcommand{\sssu}{\subsubsection}
\newcommand\ssssu[1]{\textbf{#1}}
\newcommand{\nid}{\noindent}
\newcommand{\nnb}{\nonumber}
\newcommand\bloc[2]{{\omega}_{#1}^{#2}}
\newcommand\blocp[2]{{\omega'}_{#1}^{#2}}
\newcommand{\un}{\mathbbm{1}}
\newcommand{\setZ}{\mathbbm{Z}}
\newcommand{\setN}{\mathbbm{N}}
\newcommand{\setR}{\mathbbm{R}}
\newcommand{\pare}[1]{\left(\, #1 \, \right)}
\newcommand{\scal}[2]{\langle\, #1 | #2 \, \rangle}
\newcommand{\bra}[1]{\left[\, #1 \, \right]}
\newcommand{\brac}[2]{\left[\, #1 \, | \, #2 \right]}
\newcommand{\Set}[1]{\left\{\, #1 \, \right\}}
\newcommand{\Setc}[2]{\left\{\, #1 \, ; \, #2 \right\}}
\newcommand{\Prob}[1]{P\left[\, #1 \, \right]}
\newcommand{\PT}[1]{\pi^{(T)}\left[\, #1 \, \right]}
\newcommand{\PTo}[1]{\pi^{(T)}_\omega\left[\, #1 \, \right]}
\newcommand{\pT}{\pi^{(T)}}
\newcommand{\Probex}[1]{P^{(T)}\left[\, #1 \, \right]}
\newcommand{\probc}[2]{P[#1 \, \left| \, #2 \right.]}
\newcommand{\Probc}[2]{P\bra{#1 \, \left| \, #2 \right.}}
\newcommand{\Probcex}[2]{P^{(T)}\bra{#1 \, \left| \, #2 \right.}}
\newcommand{\Probcp}[2]{P^{(ap)}\bra{#1 \, \left| \, #2 \right.}}
\newcommand{\abs}[1]{\left|\, #1 \, \right|}
\newcommand{\deq}{\stackrel {\rm def}{=}}
\newcommand{\sqsubsetneq}{\stackrel {\tiny\sqsubset}{\tiny \neq}}
\newcommand\moy[1]{\mu\bra{#1}}
\newcommand\noy[1]{\nu\bra{#1}}
\newcommand\dKL[2]{d_{\tiny{KL}}\bra{#1 \, \mid #2}}
\newcommand\cA{{\cal A}}
\renewcommand{\P}[1]{\cP\bra{#1}}
\newcommand\Minv{\mathcal{M}_{inv}}
\newcommand\cB{{\mathcal{B}}}
\newcommand\cC{{\mathcal{C}}}
\newcommand\cD{{\mathcal{D}}}
\newcommand\cT{{\mathcal{T}}}
\renewcommand\H{{\cH}}
\newcommand{\h}{\mathtt{h}}
\newcommand{\J}{\mathtt{J}}
\newcommand\s[1]{{\cS\bra{#1}}}
\newcommand\tH{{\tilde{H}}}
\renewcommand\th{{\tilde{h}}}
\newcommand\tFi{{\tilde{\Phi}}}
\newcommand\tfi{{\tilde{\phi}}}
\newcommand\tF{{\tilde{F}}}
\newcommand\tG{{\tilde{G}}}
\newcommand\tu{{\tilde{u}}}
\newcommand\tv{{\tilde{v}}}
\renewcommand{\sb}{s_\bbeta}
\newcommand\cE{{\mathcal{E}}}
\newcommand\cG{{\cal G}}
\newcommand\cH{{\mathcal{H}}}
\newcommand\cI{{\mathcal{I}}}
\newcommand\cK{{\cal K}}
\newcommand\cL{{\cal L}}
\newcommand\cM{{\mathcal{M}}}
\newcommand\cS{{\cal S}}
\newcommand\cP{{\mathcal{P}}}
\newcommand\cO{{\cal O}}
\newcommand\cU{{\cal U}}
\newcommand\cW{{\cal W}}
\newcommand\f{{\bm{f}}}
\newcommand\be{{\bm{e}}}
\newcommand\bh{{\bm{h}}}
\newcommand\bxi{{\bm{\xi}}}
\newcommand\bn{{\bm{n}}}
\newcommand\bv{{\bm{v}}}
\newcommand\bF{{\bm{F}}}
\newcommand\bo{{\bm{o}}}
\newcommand\blambda{{\bm{\lambda}}}
\newcommand\bbeta{{\bm{\beta}}}
\newcommand\gk[1]{g_k\pare{#1}}
\newcommand\gLk{g_{L,k}}
\newcommand\akj[1]{\alpha_{kj}(#1)}
\newcommand\sif[1]{\seq{\omega}{#1}{D}}
\newcommand\Xk[1]{X_k({#1})}
\newcommand\Ik[1]{I_k({#1})}
\newcommand\Skj[1]{S_{kj}({#1})}
\newcommand\X[2]{X_{#1}\pare{#2}}
\newcommand\et[2]{\eta_{#1}\pare{#2}}
\newcommand\tO[2]{\tau_{#1}\pare{#2}}
\newcommand\Vd[2]{\mathcal{V}^{(det)}_{#1}\pare{#2}}
\newcommand\XkD{\Xk{\bloc{0}{D-1}}}
\newcommand\Xkr{X_k^r}
\newcommand\tLk{\tau_{L,k}}
\newcommand\moyc[2]{\mu\bra{#1 \, | \, #2}}
\newcommand\Vkdet[1]{V_k^{(det)}({#1})}
\newcommand\Vksyn[1]{V_k^{(syn)}({#1})}
\newcommand\Vkext[1]{V_k^{(ext)}({#1})}
\newcommand\Vknoise[1]{V_k^{(noise)}({#1})}
\newcommand\sk[1]{\sigma_k({#1})}
\newcommand\skr{\sigma_k^r\pare{\bloc{0}{D-1}}}
\newcommand{\ent}[1]{\left[\, #1 \, \right]}
\newcommand\vm[1]{var_m\left[#1 \right]}
\newcommand\eg[1]{\stackrel {\rm #1}{=}}
\newcommand\tk[1]{\tau_k\pare{#1}}
\newcommand\tko{\tau_k(t,\omega)}
\newcommand\cBm[1]{\cB^{-1}\pare{#1}}
\newtheorem{theorem}{Theorem}%[CTh]
\newcommand{\bth}{\begin{theorem}}
\newcommand{\enth}{\end{theorem}}
\newcommand\omD[1]{\seq{\bra{\omega^{(#1)}}}{0}{D-1}}
\newcommand\omz[1]{\omega^{(#1)}(D)}
\newcommand\om[1]{\omega^{(#1)}}
\newcommand\Om[1]{\Omega^{(#1)}}
\newcommand\skjq{s_{kj;q}}
\newcommand\ikq{i_{k;q}}
\newcommand\skjqs{p_{kj;q_s}}
\newcommand\ikqs{i_{k;q_s}}
\newcommand\ikqsu{i_{k;q_{s_u}}}
\newcommand\ikqu{i_{k;q_u}}
\newcommand\xkq{x_{k;q}}
\newcommand\skjqv{s_{kj;q_{v}}}
\newcommand\tjr{t_j^{(r)}(\omega)}
\renewcommand\S[2]{S_{#1}\pare{#2}}
\newcommand{\im}{Im}
\newcommand\rpf{R}
\newcommand\rpfc[1]{R\pare{#1}}
\newcommand\lpf{L}
\newcommand\lpfc[1]{L\pare{#1}}
\newcommand{\zb}{\zeta_{\bbeta}}
\newcommand{\etab}{\eta_{\bbeta}}
\newcommand{\mex}{\mu^{(ex)}}
\newcommand{\phex}{\phi^{(ex)}}
\newcommand{\Phex}{\Phi^{(ex)}}
\newcommand{\Hex}{\cH^{(ex)}}
\newcommand{\hex}{h^{(ex)}}
\newcommand{\siex}{\sigma^{(ex)}}
\newcommand{\cn}{\cC^{\ast}}
\newcommand\p[1]{\cP\bra{#1}}
\newcommand\w[2]{w_{#1}^{(#2)}}
\newcommand\Gb{\cG_\bbeta}
\newcommand\K[2]{\cK_{\small{#1; \,  #2}}}
\renewcommand\d[2]{\delta_{\small{#1; \,  #2}}}
\newcommand\blp[3]{\omega^{#3,(#1)}_{#2}}
\newcommand\blpb[3]{\overline{\omega^{#3,(#1)}_{#2}}}
\newcommand\moyex[1]{\mu^{(ex)}\bra{#1}}
\newcommand\E[1]{{\cal E}_{#1}}
\newcommand{\tc}{\tau(\cC)}
\newcommand{\notsqsupset}{\cancel{\sqsupset}}
\newcommand{\sm}{\sigma^{-1}}
\newcommand\hP{\hat{\cP}}
\newcommand{\pp}{\mathbb{P}}
\newtheorem{ex}{\textbf{Example}}